# The Relationship between Cognition and Computation： "Global-first" Cognition versus Local-first Computation

Lin Chen [1]

[1]University of Chinese Academy of Sciences and Institute of Biophysics, Chinese Academy of Sciences, Beijing, China

What fundamental research questions are essential for advancing toward brain-like AI or AGI (Artificial General Intelligence) capable of performing any intellectual task a human can? Should it be something like the Turing machine (1936), which answers the question "What is computation?" and lays the foundation for the entire field of computer science? Or should it be something like Shannon's mathematical theory of communication (1948), which answers the question "What is information?" and forms the basis for modern communication technology?

We believe the key question today is the relationship between cognition and computation (RCC). For example, the widely discussed question "Will artificial intelligence replace the human mind?" is, in essence and in scientific terms, an issue concerning RCC.

We have chosen to classify RCC into four categories:

1. The relationship between the primitives of cognition and the primitives of computation.
2. The relationship between the anatomical structure of neural representation of cognition and the computational architecture of artificial intelligence.
3. The relationship between emergents in cognition and emergents in computation.
4. The relationship between the mathematical foundations of cognition and computation.

## 1. The Relationship between the Primitives of Cognition and the Primitives of Computation.

One of the most fundamental questions about any process concerns the nature of the primitive units over which it operates. Each basic science has established its own field-specific basic units—for example, symbols in computational models, bits in information theory, genes in genetics, and elementary particles in high-energy physics. In cognitive science, this fundamental question becomes: What are the primitive units over which cognitive processes operate, or, at its core, "Where to begin?" (e.g., Chen, 1982, 2005; Humphreys & Riddoch, 1987; Todd et al., 1998; Pomerantz, 2003; Wolfe, 2005; He, 2008; Todd & Petrov, 2022).

It is well worth emphasizing that ample empirical evidence accumulated across various fields and different aspects of cognitive science reveal that the primitives of cognition are *neither* symbols of computation *nor* bits of information. For example, in the field of short-term memory, Miller (1956), in his highly cited paper "The Magical Number Seven, Plus or Minus Two", demonstrated that one can repeat back a list of no more than about seven, plus or minus two, randomly ordered items or chunks (which could be letters, digits, or words). Hence, Miller persuasively argued that bits of information theory were not the basic units of short-term memory; rather, holistic *chunks* were. One chunk can be a digit, a letter, a whole word composed of multiple letters, or even a sentence composed of multiple words, which obviously carry vastly different amounts of information. More concretely, binary digits have 1 bit each, decimal digits have 3.32 bits each, and words have about 10 bits each.

In addition to *chunks* in working memory, a greater number of representative models of cognitive primitives can be illustrated in various types or aspects of cognition. These include, for example, *object identity* in motion perception, *objects* in selective attention, *temporal Gestalts* in temporal organization, and "the integrative self" in consciousness.

In the field of visual motion, Marr (1981) proposed *object identity* as the primitive of shape-changing motion. He argued, "Time introduces an important new factor, which is rather independent of the precise details of an object's three-dimensional structure. This factor is the consistency of an object's identity through time, and it is a different problem entirely". In Marr's book *Vision*, *object identity* was proposed in the particular context of visual motion. However, from our perspective, the concept of *object identity*, as the primitive of shape-changing motion, is a farther-reaching idea for understanding visual cognition in general. It is unfortunate that, while Marr's well-known primal sketch and edge-detector model serve as the main theme of his book *Vision*, his idea of object identity has been mostly overlooked in contemporary literature.

In the field of attention, regarding what the basic units selected by attention are, Kahneman and Henik (1981) claimed that the primitives of selective attention are holistic *objects*, which are formed by preliminary perceptual organization. While the majority of theories assume that attention selects some featural properties, such as spatial location, the object-based theory states that to the degree a perceptual object is attended, all responses associated with the properties or components of the object will be facilitated, and particularly, emphasized that "the object is prior to its properties and independent of them".

In the field of temporal organization, Pöppel (1997) claimed that the primitives of temporal perception are *temporal Gestalts*. He illustrated his point by explaining that the very notion of "now" or "present" depends on an integrative mechanism that fuses chronologically sequential events into "a present Gestalt". For example, when we read or hear the word "now", we perceive the whole "now" rather than the sequence of three different speech sounds, n-o-w, which are fused into a holistic unit of temporal organization. This analysis indicates that while 'the present' or simultaneity is largely considered to have zero duration in the physical world, "a present Gestalt" has a flexible temporal extent (up to 2.5-3 seconds as observed). Obviously, this fundamentally and particularly constrains all mental (rather than physical) states and processes.

In the domain of the "self as object", Sui and Humphreys (2015) argued for "the integrative self" as a primitive of consciousness. They measured the self using an associative-matching procedure, where participants were instructed to associate a personal label (e.g., you, friend, or stranger) with a geometric shape (e.g., triangle, circle, or square). Participants then completed a shape-label matching task where shape-label pairs were presented either in their original assignment or not. Typically, results showed more accurate and faster responses to self-associated shapes compared to other-associated shapes. Based on the discovery of the "self-prioritization effect", they argued for "the integrative self" in the context that self-reference notably affects the integration of parts into perceptual wholes and acts as a form of integrative hub or associative "glue" to integrate different types and stages of cognitive processing. It turns out that holistic operations in perceptual organizationsuch as grouping, belonging, and determining "what goes with what"—which are central to the notions of objects in attention and chunks in working memory, are also fundamental to the nature of "the integrative self" as a primitive of consciousness. This holds true even though the self seems to involve very different types of cognitive processes compared to selective attention and working memory.

Regarding these primitives of cognition discussed above, the present paper emphasizes their unique role in understanding RCC. Specifically, concepts such as *chunks*, *objects* and *object identity*, *temporal Gestalts*, and *"the integrative self"* are not borrowed from information theory, computer science, physical sciences, or other fields of science. Rather, they originated from cognitive science itself, particularly from its experimental findings and empirical observations. Thus, they hold an important and irreplaceable position in our endeavor to define the primitives of cognition.

A major challenge we now face is how to define these primitives of cognition in a formal and unified manner, going beyond intuitive understanding. This definition is of fundamental importance for understanding RCC and serves as a critical step toward making artificial intelligence a mature basic science, akin to the transition "from alchemy to chemistry". To meet this challenge, a "Global-first"* principle will be introduced below.

**The "Global-first" Principle**

To address RCC, our efforts over more than four decades have led us to formulate the "Global-first" principle. By "Global-first," we declare, in so many words:

- *To grasp the very notion of "global", the "Global-first" principle defines "global" specifically as topological, providing a mathematically grounded definition.*
- *Supported by empirical evidence as well as formal analysis, perceptual organization based on global topological invariants serves as the starting point as well as the basic structure of cognitive processes.*
- *In particular, with respect to both time dependence and logical dependence, the global topological property—based on spatial and/or temporal proximity—is prior to, and independent of, local geometrical and other featural properties.*
- *The "global-first" principle is also embodied in Zeeman's tolerance spaces and homology theory—a branch of mathematics well-suited to ignoring local variations and capturing global properties. Since perceptual organization is essentially discrete, a critical challenge in formalizing both its mechanism in particular, and the global topological approach in general, is how to define global properties in discrete sets. Zeeman's homology theory of tolerance spaces provides a solution by defining global properties in discrete sets in terms of global tolerance invariants specified by homology groups in tolerance spaces. In general, by their very nature, tolerance spaces render any local topological structure irrelevant; on the other hand, the global topological structure turns out to be highly relevant. This intrinsic—and seldom seen in most branches of mathematics—global-first structure makes tolerance spaces a mathematical foundation for the "global-first" principle.*
- *In addition, the "global-first" principle is reflected in a perceptual hierarchy, the relative perceptual salience of geometrical invariants. This hierarchy systematically corresponds to the structural stability of these invariants under transformation, in a manner similar to Klein's Erlangen hierarchy of geometries—specifically, in a descending order from global to local, with topological invariants being "global-first," followed by projective, affine, and Euclidean invariants.*

A major contribution made by the "Global-first" principle is the establishment of the "Global-first" topological definition of the primitives of cognition. This definition claims that:

*Primitive units of cognitive processing—including, for example, object identity in motion, objects in attention, chunks in working memory, temporal Gestalts, and "the integrative self"—can be understood as entities that maintain their topological structure over time. Their common and core intuitive notion—the global identity that emerges and is preserved through holistic operations in perceptual organization, such as shape-changing transformations, grouping, belonging, and "what goes with what"—can be characterized in a formal and unified manner by topological invariants of connectivity, the number of holes, and the inside/outside relationship.*

A basic but counter-intuitive inference drawn from this topological definition is that the topological change of an object should consequently disturb its object continuity and be perceived as the emergence of a new object. Conversely, object identity survives non-topological changes of local as well as physical features.

The "Global-first" topological definition of the primitives of cognition has been rigorously tested and verified across a wide variety of cognitive processes, encompassing multiple topics and fields within cognitive science (see Note for a brief description of each): 1. Visual sensitivity and discriminability (Note 1.); 2. The relationship between Gestalt principles (Note 2.); 3. The figure-background relationship (Note 3.); 4. The perception of shape invariance (Note 4.); 5. Immediate and effortless texture segregation (Note 5.); 6. Defining the concept of perceptual objects: cognitively real and formally general (Note 6.); 7. Temporal organization (Note 7.); 8. Configural superiority effects (Note 8.); 9. The object-superiority effect (Note 9.); 10. Global precedence (Note 10.); 11. The redundant-target effect (RTE) (Note 11.); 12. Apparent motion and its correspondence token (Note 12.); 13.Visual search (Note 13.); 14. Attentive tracking (Note 14.); 15. Precuing attention (Note 15.); 16. Attentional captures (Note 16.); 17. Attentional blink (Note 17.); 18. The crowding effect (Note 18.); 19. Eye movement (Note 19.); 20. Peripheral vision (Note 20.); 21. Form factors in visual masking (Note 21.); 22. Illusory conjunctions (Note 22.); 23. Stereoscopic 3D-form discrimination (Note 23.); 24. Working memory capacity (Note 24.); 25. Perceptual learning and its transferability (Note 25.); 26. Asymmetry of hemispheres (Note 26.); 27. Numerosity perception (Note 27.); 28. Perceived animacy (Note 28.); 29. Olfactory sense (Note 29.); 30. Interaction between fear and topological discrimination (Note 30.); 31. Binocular rivalry (Note 31.); 32. Visual illusions and magical experiences (Note 32.); 33. Self as an object (Note 33.); 34. Infants' and children's visual cognition (Note 34.-35.); 35. Visual pattern recognition in animals (Note 36.-40.); 36. ipRGCs (intrinsically photosensitive retinal ganglion cells) vision (Note 41.); 37. Epilepshy (Note 42.); 38. Hemispatial neglect (Note 43.); 39. Myopia (Note 44.); 40. AD (Alzheimer's disease) and MCI (Mild cognitive impairment) (Note 45.); 41. Williams syndrome (Note 46.); 42. Autism spectrum disorder (Note 47.); 43. Schizophrenia (Note 48.); 44. Blindsight (Note 49.); 45. Aging vision (Note 50.); 46. Subcortical pathways (Note 51.); 47. fMRI scan of apparent motion (Note 52.); 48. The baseline of brain function in the context of fMRI (Note 53.); 49. Computational analysis and modeling (Note 54.-55.); and others (Note 56.-59.).

The "Global-first" topological definition of the primitives of cognition formally and deeply contrasts with the primitives of computation, such as bits and symbols, and offers a well-defined and achievable scientific proposition. This proposition may lead to a breakthrough in understanding the relationship between the primitives of cognition and the primitives of computation in particular, and RCC in general.

*The term "Global-first" was coined by Jeremy Wolfe.

## 2. The Relationship between the Anatomical Structure of Neural Representations of Cognition and the Computational Architecture of Artificial Intelligence.

When discussing the architecture of the brain, two main views stand at opposite poles: modularity and connectionism. These concepts are so fundamentally different that they appear to be literal opposites. Although issues in brain science and cognitive science are far more complex than this dichotomy suggests, such basic polarization can serve as a useful probe to estimate the current level of our knowledge about brain structure. If it cannot yet be determined whether the brain's basic architecture adheres to modularity or, conversely, to connectionism, caution should be exercised when discussing brain-like computation. This is because we do not yet have a clear and precise understanding of what "brain-like" truly means.

A few comments on modularity versus connectionism are provided below. Firstly, we will discuss modularity in relation to brain mapping. Modularity claims that the brain consists of many modules, each performing separable cognitive functions autonomously from other modules. When conducting brain mapping, we largely assume a modular structure of the brain and attempt to localize which brain regions perform certain cognitive functions, separate from other regions. The advent of brain imaging techniques, such as fMRI, has had a dramatic impact on this field. However, it is the concept of modularity that serves as the underlying driving force behind the rapid growth of human brain mapping. Modularity in relation to brain mapping highlights an obvious distinction between the modular architecture of the brain and the CPU-central storage architecture of general-purpose computers. This distinction suggests, from both philosophical and empirical perspectives, the abandonment of the once widely held metaphor that the brain is akin to a digital general-purpose computer in architecture.

Despite the apparent success of modularity in brain mapping, its limitations can be illustrated through one of our recent findings, the Topological Discrimination Network (TDN) (see Figure 1). The TDN is a cortical network defined by task-evoked deactivations. It is identified by contrasting the topological discrimination task with local geometric discrimination tasks, such as orientation or mirror-symmetry discrimination. Interestingly, the magnitudes of TDN deactivations in each of the TDN hubs were consistently modulated by, for example, orientation discrimination (Figure 1B). Specifically, significantly greater magnitudes of TDN deactivations were found for orientation discrimination than for topological discrimination. This consistent modulation by orientation indicates that the perception of orientation involves the more than 13 TDN hubs at the whole-brain scale, including the subcortical pathway, rather than being localized to a single (e.g., V1) or a few neighboring brain regions. Historically, localizing orientation detectors in line segments was popular and seemingly successful in neurophysiology, significantly impacting both symbolic and connectionist modeling of vision. However, the discovery of the TDN demonstrates the limitations of such a localization approach and does not favor modularity. It suggests that while brain mapping is a valuable approach to understanding brain architecture, it should go beyond mere region-localizing brain mapping.

Finally, let's discuss the central idea of connectionism. As the term implies, connectionism posits that the connections linking cells determine the function of neural networks, while the inner structures of cells are considered of no great importance. However, a recent surprising discovery challenges this view, suggesting that topological perception originates in a particular kind of neuron, namely, intrinsic photosensitive retinal ganglion cells (ipRGCs) in mice (see Note [157]). Mice that retain ipRGCs but lack rod and cone phototransduction (called MO mice) are largely considered functionally blind and non-image forming. However, the "Global-first" principle led to a counter-intuitive hypothesis that MO mice, despite being blind to local featural properties, might still preserve the capacity for topological perception. Otherwise, meaningful vision in MO mice would be impossible, as the global organization based on topology serves as the starting point of cognition. To test this hypothesis, the ability of MO mice and other transgenic mice to perceive topological differences was examined using various behavioral paradigms, such as fear conditioning and water maze tests. The results consistently demonstrated that MO mice can indeed discriminate topological differences but not non-topological form differences. MO mice learned to discriminate stimuli based on topology, abstracted the topological invariants of the stimuli, and transferred their learned topological discriminability to novel stimuli. Particularly convincing are the results from improved control conditions using gene manipulation techniques. Rescuing melanopsin in ipRGCs of TKO (total knockout) mice restored their capability for topological discrimination. In contrast, melanopsin-free mice (OPN4-KO), even with intact rods and cones, were unable to consistently discriminate topological properties. These findings strengthen the conclusion that ipRGCs are necessary and sufficient for consistently discriminating topological differences. Moreover, only two or three of the five subtypes of ipRGCs may be involved in image-forming and possibly in topological perception. These results collectively suggest that the "Global-first" topological perception originates in a specific type of cell with particular morphology, projection, and expression profile. In contrast, cells in artificial neural networks are treated as having no inner structure and in an undifferentiated manner. In summary, this experimentally well-controlled and mathematically well-defined discovery highlights a fundamental limitation of connectionism: while connections are important, the inner structure of neurons is too.

To sum up, if deep learning were considered to be intrinsically inspired by the hierarchical structure of neural systems, then brain mappings that go beyond mere region localization—such as TDN mapping at the whole-brain scale and ipRGCs mapping at the neuronal and molecular levels—would offer much more essential and profound inspiration for developing new computational model architectures.

## 3. The Relationship between Emergents in Cognition and Emergents in Computation.

When discussing the far-reaching influence of emergence on cognitive sciences and AI, McClelland (2010) argued that "The greatest achievements of human cognition may be largely emergent phenomena" and "It remains a challenge for the future to learn more about how these greatest achievements arise and to emulate them in artificial systems".

Emergence commonly refers to holistic phenomena in complex systems that are not present in their individual parts. According to this common understanding, emergent phenomena are ubiquitous, appearing in everyday life, various scientific fields, and different branches of mathematics and philosophy. This paper, however, focuses on emergents in cognitive constructs, particularly on the primitives of cognition as emergent phenomena. When discussing examples of primitives of cognition such as *objects*, *chunks*, *temporal gestalts*, and "the integrative object", the topological definition highlights "their common and core intuitive notion—the global identity emerged and preserved over holistic operations". In this context, these examples of cognitive primitives are essentially viewed as emergents.

**Reformulating Classic Gestalt Holism in Understanding Emergence: The Global is *Prior* to the Local.**

To understand RCC from the perspective of emergence, we appeal to Gestalt theory. Historically well-known, Gestalt theory is particularly illuminating in exploring the whole/part or global/local relations. The famous phrase "the whole is more than the sum of its parts", commonly quoted as the core belief of Gestalt psychology, has been used to define emergence. However, this phrase is actually an incorrect translation of Koffka's original phrase (1935), "the whole is something else than the sum of its parts". Koffka, a founder of Gestalt psychology, did not favor the translation using "more" and firmly corrected his students who replaced "other" with "more". He emphasized, "This is not a principle of addition".

In our view, the original value of Gestalt holism with respect to emergence lies in its assertion that the whole is not only more than, but also not merely "very different from" (as emphasized by Anderson in his landmark paper "More Is Different") the sum of its parts; rather, the whole is independent of its parts and possesses a reality of its own.

However, the "Global-first" principle, transcending Gestalt holism, reformulates it as: "The global is *prior* to the local". As discussed below, the emphasis on *prior*—rather than "more" or "different"—is central to identifying the "Global-first" characteristic of emergent cognitive primitives.

In the "Global-first" principle, the reformulation of "The global is *prior* to the local" replaces various classic Gestalt expressions such as "The whole is *more* than the sum of its parts", "The whole is *greater* than the sum of its parts", and "The whole is *different* from the sum of its parts". Here, the emphasis on *prior*—rooted in the 'Global-first' principle—distinguishes it from the classic Gestalt focus on "more," "greater", or "different".

*The True Value of Prior in Understanding Emergence*

The keyword *prior* is repeatedly emphasized here because it may be crucial in distinguishing between the emergence of cognitive constructs and computational constructs. While most researchers readily accept the concept of emergence as "the whole is more than the sum of its parts," this formulation applies similarly to both cognitive and computational emergence when referring to "more" or "different". However, the question of "Where to begin?"—global-to-local or local-to-global—introduces the notion of *prior*, which brings in the fundamentally new dimensions of temporal and logical dependence. These dependencies, in turn, make the relationship between emergents in cognition and in computation an entirely different problem.

In the context of "Global-first", *prior* strictly means that, in both time dependence and logical dependence, the global topological property—defined by spatial and/or temporal proximity—is perceived early and directly. Here, "early" refers to perception that occurs before local geometric features are recognized, and "directly" means that this perception takes place without first processing local geometric properties. In fact, local geometric perception depends on the preceding global topological perception; thus, measures of local geometric properties are contingent upon the global organization established by topology.

While emergence in computational constructs generally adheres to the local-first approach, emergence in cognitive constructs—particularly cognitive primitives—exhibits the "Global-first" character. It is the emphasis on the keyword *prior* in the reformulation of Gestalt holism that highlights the essence of "Global-first" in cognitive emergence, setting it apart from the predominantly local-first nature of computational constructs.

*The "Global-first": One of the Greatest Achievements of Human Cognition as Emergence*

The "local-first" line of thinking, or the approach of going "from local to global", is often taken for granted as it aligns well with the naïve and intuitive view of physical items, where components are naturally and readily seen as prior to the compound they form. The "local-first" idea is also well-supported by overwhelming biological evidence as well as most mathematical constructs. However, emergent phenomena in cognitive science offer an opportunity to challenge this dominant local-first perspective. Cognitive emergents, particularly the primitives of cognition as achievements of the mind or mental world, may not necessarily follow the physical logics and complexity analyses of computation.

A fundamental position of the "Global-first" principle is that we cannot simply determine, based on physical logics or mathematical analyses of complexity, whether a variable suitable for a primitive of physical or mathematical constructs is also suitable for a primitive of cognitive constructs in the mental world. That is, a variable which may be suitable for a primitive of physical or computational constructs does not necessarily mean it is also suitable for a primitive of cognition. It may be worth repeatedly emphasizing that physically simple or computationally primitive does not necessarily mean psychologically simple or perceptually primitive. The true value of "Global-first" lies in the fact that it does not primarily come from logical reasoning applied in the physical world or mathematical analyses of computational complexity, but rather it emerges from empirical investigation in the mental world.

The "Global-first" principle may initially seem counterintuitive in the context of the physical world and contrary to common expectations in the context of mathematical constructs. However, this impression of being counterintuitive and contradictory may simply reflect the emergent nature of the "Global-first" principle, which emerges in the cognitive or mental world. While the "local-first" approach is largely based on the consideration of emergence in most physical and mathematical systems, the "Global-first" principle conveys a unique and irreplaceable nature of emergence in cognition. In this sense, the "Global-first" principle itself should be considered a well-representative example of "the greatest achievements of human cognition" (McClelland, 2010) as emergent phenomena.

## 4. The Relationship between the Mathematical Foundations of Cognition and Computation.

From the perspective of either the classic computation model or the popular Bayesian theory, cognitive processes have often been largely modeled as symbolic manipulation akin to that of the Turing machine. Clearly, the Turing machine provides the mathematical foundation for computation.

However, computation-based mathematics do not seem to satisfactorily characterize many basic cognitive phenomena and issues, including the "Global-first" approach, the primitives of cognition, direct perception of invariance, mental imagery, metaphor of semantics, the emergence of consciousness, cognitive bias, and much more. Searching for new mathematical frameworks suitable for describing "Global-first" cognition is essential for understanding RCC. In this regard, a basic question arises: How can we define "global" in a way that is both cognitively real and mathematically grounded?

*Defining "Global" to be Cognitively Real and Mathematically Grounded*

The global versus local (or whole versus part) relationship has been a central and long-standing concern, from Gestalt psychology to contemporary cognitive sciences and artificial intelligence. Additionally, in modern theoretical physics and mathematics, Yang (1997) argued: "The relationship between the local and the global has been elevated to 'a prominent and highly regarded field' (the term '显学', xiǎn xué, cited from Chinese classic literature) in mathematics through the development of topology, Lie groups, and differential geometry in the 20th century".

However, in the literature, the global versus local relationship is linked to numerous conceptual dichotomies, such as "whole versus part", "compound versus component", "large versus small", "finer versus coarser granularity", "low versus high spatial frequencies", "local texture element versus global (statistical)", "global versus local motion", "holistic versus analytic", "integral versus separable", "undifferentiated versus differentiated", "integral versus nonintegral", and "nonanalyzable versus analyzable", among many others. This wide variety of dichotomies indicates a lack of clear understanding of what "global" exactly means. To clarify this conceptual confusion, the topological approach applies the "global versus local" distinction instead of this variety of dichotomies, such as the commonly used whole versus part, compound versus component, large versus small, finer versus coarser granularity, and low versus high spatial frequencies. Defining the concept of "global" in a cognitively real and formally grounded manner remains a major challenge to establishing a formal description of cognitive organization in general and to formulating the "Global-first" principle in particular.

We believe that certain branches of mathematics may provide insights into how to understand the concept of "global" with formal precision and psychological validity. Here, three mathematical branches are highlighted to serve as the foundations for the "Global-first" principle: *Perceptrons* by Minsky, Tolerance spaces and homology by Zeeman, and Klein's Erlangen Program.

## *Perceptrons* by Minsky as the Mathematical Foundation for Analyzing the Complexity of Computing Topological Properties.

We believe that one of the main contributions of *Perceptrons* (Minsky & Papert, 1988) is defining the complexity (order as a measure of complexity) of computing, specifically, geometrical invariants in neural networks. One of the most influential theorems in *Perceptrons* is:

"**Theorem 5.9:** The only topologically invariant predicates of finite order are functions of the Euler number *E(X)*."

This theorem reveals the limitation of perceptrons in computing topological invariants. Specifically, for perceptrons, all topological predicates, except for the predicate of the Euler number *E(X)*, are not of finite order. However, low-order perceptrons can compute local geometrical invariants.

*"Global-first" Topological Perception: A Current Century Cloud Over Computational Approaches Based on Local Features.\**

*Perceptrons*, therefore, laid down the mathematical foundation for analyzing the contradiction between the "Global-first" perception and the high computational complexity (non-finite order) of topological invariants. On one hand, from the perspective of computational complexity analysis, the global nature of topological properties makes their computation difficult. On the other hand, empirical data consistently support the "Global-first" principle, claiming that the extraction of topological properties serves as the starting point of cognition. The order of degrees of difficulty in computing geometrical invariants is puzzlingly reversed compared to the order of priority in perceiving them (Chen, 1989). This deep contradiction suggests that the global nature of topological invariants is associated with high complexity (non-finite order) as defined by *Perceptrons*.

Historically, the limitation of perceptrons in computing topological invariants has been a core issue in the debate between traditional AI and connectionism. The development of Backpropagation led to a notable argument in the PDP book, asserting that the limitation to topological computation identified by *Perceptrons* applies only to simple one-layer perceptrons and not to multilayer perceptrons. In fact, this argument provided a theoretical basis for the resurgence of connectionism in the 1980s. However, in our opinion, since *Perceptrons*' result concerns complexity rather than computability, it is insufficient to question its validity and generality solely by developing examples of networks capable of computing topological properties under certain limited conditions. It might even be possible to construct some kinds of multilayer networks that can compute certain aspects of topological structure in real time under specific conditions. Nevertheless, in comparison with local geometrical properties, topological properties still exhibit high complexity, even for multilayer networks (Chen, 2005, pp. 629-633 and Note 55.). Minsky and Papert (1988) argued that multilayer networks with backpropagation could compute some topological properties, but as the sample size of the input increases, the number of hidden units needed and the time required to learn will increase tremendously. "Given this consideration and the fact that no success has been reported in recognizing connectedness, it is reasonable to project, based on the result for simple perceptrons, that the limitation exists for multilayer perceptrons" (Wang, 2000).

In short, *Perceptrons*' classic results on the high complexity of topological invariants are, in our view, mathematically valid in general. As a consequence, the long-standing challenge remains for computational approaches based on local geometrical features to explain the "Global-first" topological perception. The "Global-first" topological perception, therefore, inevitably raises fundamental and challenging issues about the relationship between topological cognition and topological computation in particular, and the relationship between "Global-first" cognition and local-first computation in general (Chen, 2005, pp. 629-633 and Note 55.).

\* This title is in the spirit of the famous historical metaphor of clouds in theoretical physics from Lord Kelvin's "Nineteenth Century Clouds Over the Dynamical Theory of Heat and Light" (1901).

**Zeeman's Tolerance Spaces and Homology as the Mathematical Foundation of the "Global-first" Perceptual Organization.**

Visual processing is essentially discrete, from image sampling by the cone and rod mosaic to gestalt-like grouping of discrete stimuli such as disconnected dot arrays. However, our subjective perception and experience are fundamentally continuous and holistic. This contradiction indicates that general (continuous) or point-set topology is not readily applicable to describe perceptual organization. A basic and critical issue for the "Global-first" principle is how to mathematically define global properties in a discrete set. Specifically, what attributes of a physically disconnected stimulus (such as a dot array) determine its perceptual or subjective connectivity? To address this issue, we need a special branch of topology that, like general topology, is adept at capturing global properties such as connectivity and holes, while also being well-suited for application to discrete sets.

Tolerance spaces, as introduced by Zeeman (1962, 1968), are considered well-suited for capturing global properties while ignoring local details within a certain tolerance. Thus, they are applied to establish the mathematical foundation of the "Global-first" perceptual organization, particularly in defining global properties in discrete sets as global properties of homology in tolerance spaces.

A tolerance is defined in terms of an algebraic relation that is reflexive and symmetric, but generally not transitive. This intransitivity gives tolerance spaces special mathematical properties, with a strong emphasis on global properties. While differential equations are suitable for describing changes in local properties, the global tolerance structure is more appropriate for addressing global properties. As Zeeman (1968) argued, "By the very nature of a tolerance space, any local topological structure is irrelevant. On the other hand, the global topological structure turns out to be highly relevant".

In a tolerance space, we can build mathematical structures such as homology to handle global properties. Zeeman proved that global topological properties, including tolerance connectivity, the number of tolerance holes, and dimension, are preserved under tolerance homeomorphism. This theorem provides the mathematical basis for defining global properties in a discrete set. Given a tolerance space $(X, \xi)$, we can construct a simplicial complex consisting of all simplexes, where a simplex is a finite oriented subset of X, all of whose points are within a given tolerance. The homology group $H(X, \xi)$ refers to the homology group of this complex. From $H(X, \xi)$, we can capture the global tolerance properties of the tolerance space all at once. Specifically, from the homology group $H(X, \xi)$ of a tolerance space $(X, \xi)$, we can determine how many tolerance-connected components the set X with tolerance $\xi$ consists of and how many tolerance holes each of these connected components contains. This mathematical description of global properties in a discrete set aligns well with our holistic perception. While it may seem straightforward, it is by no means trivial. It indicates that the homology theory of tolerance spaces is precisely the mathematics needed to grasp the nature of global properties in perceptual organization.

*Tolerance, in the Spirit of the "Global-first" Principle, Implies Indiscrimination beyond Mere Indistinguishability.*

The original implication of a tolerance by Zeeman corresponds to the notion of the least noticeable difference, meaning "indistinguishable", in physical or ecological world. However, the concept of tolerance used in the "Global-first" approach extends beyond the limitation of the least noticeable difference or being "indistinguishable". Instead, it essentially implies "indiscrimination" (e.g., Chen, 2005). In this context, a tolerance refers to the range within which detailed variations (which may still be distinguishable) are intentionally ignored by cognitive systems to emphasize global properties. Extending the implication of tolerance from the notion of the least noticeable difference (indistinguishable) to the notion of indiscrimination plays an important role in making the mathematical structure of tolerance spaces psychologically and ecologically relevant. At first glance, a tolerance might appear to be a capacity limitation. However, it is actually one of the most fundamental strengths and basic strategies of cognitive systems, allowing them to optimally utilize their resources. Detailed structures or changes within the tolerance are treated with indiscrimination to capture global properties under the condition of a tolerance, serving as an ecological minimum measure (Gibson, 1979).

In short, we particularly value the branch of tolerance spaces and homology, as it is rarely an inherently "Global-first" mathematics. Unlike most branches of mathematics that are built upon local-first structures, such as distance metrics, the branch of tolerance spaces and homology, which are built upon the tolerance relation rather than local-first structures like distance metrics, prominently features the global nature of being "well adapted to ignore local variations and capture global properties" (Zeeman, 1965). This global nature makes tolerance spaces and their homology particularly well-suited to serve as the mathematical foundation for the "Global-first" principle.

**Klein's Erlangen Program as the Mathematical Foundation of the "Global-first" Direct Perception of Invariance**.

The Erlangen Program, introduced by Klein (1872), provided a hierarchical structure for various geometries, including Euclidean, affine, projective geometry, and topology. These geometries are classified according to their transformation groups and the properties that remain invariant under these transformations. The impacts of the Erlangen Program extend beyond geometry, influencing various branches of mathematics and even theoretical physics. It continues to be a foundational concept in modern mathematical thought.

One of the highlights of the "Global-first" principle is the establishment of a perceptual hierarchy of the relative salience of geometrical invariants. This hierarchy is systematically related to their structural stability in a manner similar to the hierarchy of geometries in Klein's Erlangen Program. In descending order of stability (from global to local), this hierarchy of relative salience ranges from topological to projective, affine, and Euclidean invariants.

Klein's Erlangen Program and this perceptual hierarchy together led to a novel idea on how to define "global": A property is considered more global the more general the transformation group is, under which this property remains invariant. Topological transformations are the most general, and hence, topological properties are considered the most global.

To sum up, the mathematical analyses stem from three quite different branches or domains of *Perceptrons*, the homology of tolerance spaces, and Klein's Erlangen Program. Nevertheless, they all convergently point to the global nature of topological invariants. This convergence has led to the formal definition of the concept of "global" in terms of "topological" within the context of the "Global-first" principle.

**An "Erlangen Programme" for AI**

Recently, an "Erlangen Programme" for AI has been advanced by Bronstein et al. (2021) and is currently being run by a consortium of six top universities and thirteen leading British international industry and public sector partners. This AI research program, a geometric unification endeavor in the spirit of Klein's Erlangen Programme of invariance and symmetry, aims to develop a framework called "Geometric Deep Learning" to elucidate the architectures and relationships between various popular neural networks, such as CNNs, Graph Neural Networks, Transformers, and Long Short-Term Memory networks. It seeks to establish a common mathematical foundation for intelligence that focuses on the inherent symmetries and invariances in modern AI/ML systems. Every neural network architecture can be derived from fundamental principles of symmetry, in the same way that Noether's Theorem guides the development of modern physics from their corresponding symmetries. This remarkable application of Klein's Erlangen Programme in the domains of AI and ML demonstrates its profound and far-reaching influence across long historical periods and a broad range of fields, spanning historically distinct branches of mathematics and modern physics to contemporary AI.

**An "Erlangen Programme" for Visual Cognition**

Forty years before the "Erlangen Programme" for AI, an "Erlangen Programme" for visual cognition—a perceptual hierarchy of the relative salience of geometric invariants, remarkably consistent with Klein's Erlangen Programme—was already discovered in the "Global-first" principle (Chen, 1982, 2005; and Note 4.).

Besides its historical endurance, the "Global-first" principle particularly features topological invariance as its fundamental and core concern. Topological invariance, the most basic invariant in Klein's Erlangen Programme, remains largely unrecognized or ignored in most contemporary AI models. This unfortunate oversight is likely due to theoretical difficulties related to the computational complexity of topological invariants, as proved in *Perceptrons* (Minsky and Papert, 1988), and misinterpretations of empirical phenomena, such as the well-known spirals, which are, in fact, a type of ambiguous figure in connectedness caused by competing organizations of proximity and similarity (e.g., Chen, 2005, pp. 630-631). However, from our perspective, the foundation built on topological invariance is fundamentally essential for any complete theory of cognition. This is because topological invariance addresses crucial issues such as "Where to begin?", defining "global", the global versus local relationship, the primitives of cognition, and more generally, the underpinning structure of cognition. In the "Global-first" principle, the claim that "*the perceptual organization based on global topological invariants serves as the starting point as well as the basic structure of cognitive processes*" implies that topological perception is indispensable. More specifically, without topological perception, local geometrical perception would not occur. These issues are obviously central to understanding RCC in particular, and to any theory of cognition in general.

Moreover, the "Erlangen Programme" for visual cognition is supported by extensive and systematic experimental evidence, extending far beyond mere formal and mathematical models (see Section 1 and Note 4).

In summary, with its historical endurance, topological core basis, and empirical support, the "Global-first" principle, inspired by the spirit of the Erlangen Programme, extraordinarily demonstrates the broad, profound, and far-reaching impacts of Klein's Erlangen Programme.

**An "Erlangen Program" for the "Global-first" Brain Mapping**

Our recent fMRI research has further demonstrated the profound impact of the Erlangen Program through the discovery of the neural representation of the "Global-first" principle, specifically the TDN (Topological Discrimination Network) mentioned above.

A series of stimulus arrays was systematically extended to represent the different levels of geometric invariants, as stratified by Klein's Erlangen Program—including topological, projective, affine, and Euclidean invariants (see Figure 1A). This series of stimulus arrays, from left to right, represent discrimination tasks based on the following: the difference in orientation (a Euclidean invariant), the difference in parallelism (an affine invariant), the difference in collinearity (a projective invariant), the topological difference in one hole, the topological difference in two holes, and a baseline task, which is considered the easiest discrimination task of detecting "something versus nothing". In short, these discrimination stimuli constitute a hierarchy of geometries according to Klein's Erlangen Program.

This series of stimulus arrays in a texture-segregation paradigm was originally used to explore—and actually led to the discovery of—the "Erlangen Program" for visual cognition, as mentioned above (e.g., Chen, 2005). In turn, the "Erlangen Program" for visual cognition, in the context of behavioral investigation, inspired the use of fMRI scans to examine brain responses to this hierarchy of geometric invariants, in comparison to the resting state (fixation) (Note [158] and [174]). The results remarkably revealed a hierarchy of relative magnitudes of task-induced deactivations at each of the 13 TDN hubs, modulated by the geometric invariants of the Erlangen Program. As shown in Figure 1(B), at each of the 13 TDN hubs, the magnitudes of TDN deactivations increased monotonically with the global to local levels of geometric invariants of the Erlangen Program, ranging from topological, projective, and affine to Euclidean invariants, as represented by the series of stimulus arrays. At almost each of the TDN hubs, Page's trend test consistently indicated a strongly significant ordered trend among the magnitudes of TDN deactivations modulated by the geometrical invariants of the Erlangen Program. This includes the left and right IPL (Page's L = 887 and 866, respectively; for both, $p < .0001$), the vmPFC (Page's L = 880, $p < .0001$), the dmPFC (Page's L = 870, $p < .0001$), the left and right ITL (Page's L = 882 and 878, respectively; for both, $p < .0001$), the left and right PHC (Page's L = 891 and 874, respectively; for both, $p < .0001$), the left and right ATL (Page's L = 876 and 900, respectively; for both, $p < .0001$), and the left and right AMG (Page's L = 864 and 877, respectively; for both, $p < .0001$). Additionally, a significant ordered trend was also observed at the PCC (Page's L = 832, $p < .05$).

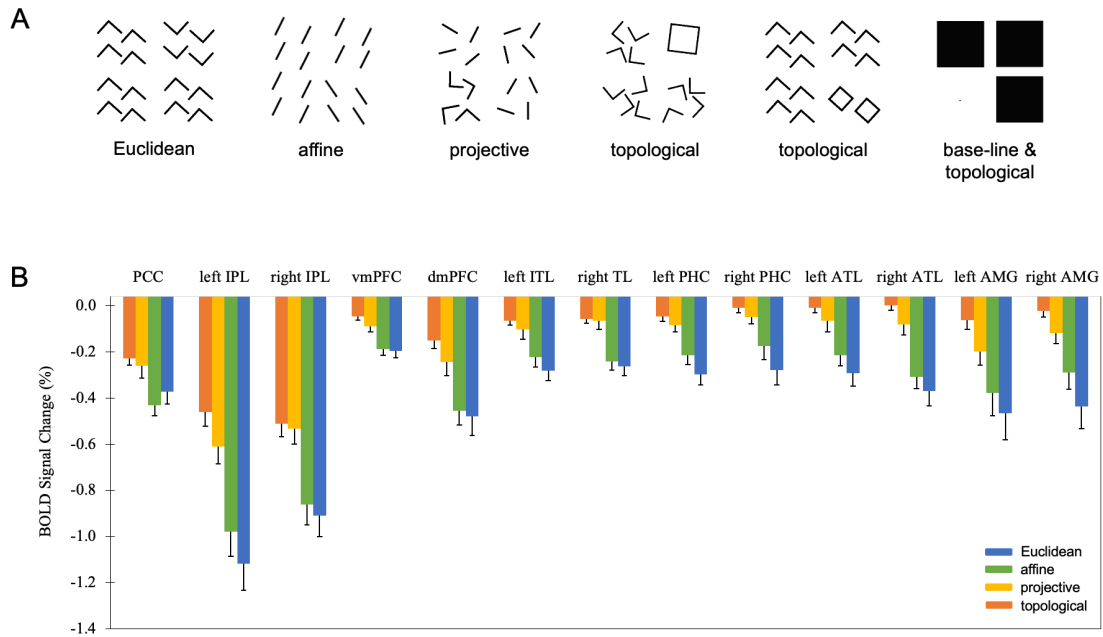

**Figure 1. An "Erlangen Program" for the "Global-first" brain mapping: a hierarchy of relative magnitudes of TDN deactivations modulated by geometric invariants.** A group BOLD activation map, contrasting the topological discrimination with the local geometrical discrimination, was generated with 169 healthy subjects. Thirteen hubs were identified (FWE-corrected (p < 0.001, ⩾ 10 voxels) as TDN ROIs. The obtained coordinates of the 13 TDN ROIs in MNI space are listed below, along with their corresponding brain bubs: (-2, -38, 38) for the PCC (posterior cingulate cortex), (-50, -70, 42) / (54, -68, 40) for the left / right IPL (inferior parietal lobule), (8, 30, -6) for the vmPFC (ventral medial prefrontal cortex), (-12, 66, 14) for the dmPFC (dorsal medial prefrontal cortex), (-56, 2, -16) / (54, 6, -24) for the left / right ITL (inferior temporal lobe), (-24, -14, -18) / (24, -4, -18) for the left / right PHC (parahippocampal cortex), (-50, 14, -28) / (50, 10, -34) for the left / right ATL (anterior temporal lobe), and (-20, 0, -22) / (24, 0, -22) for the left / right AMG (amygdala). Further detailed description can be found in the text and Note [158] and [174].

## TDN, in the Spirit of the Erlangen Program, Better Informs the Brain's Functional Baseline

*The Apparent Consistency of the TDN with the DMN*

It is rather surprising to observe the phenomenon of "default but not rest" (Note [174]). Despite clear differences in their modes of identification, the TDN—defined by task-based fMRI deactivation—and the DMN—defined by synchronous fluctuations in resting-state fMRI—are composed of essentially the same set of brain regions. Moreover, the fMRI responses underlying topological discrimination within TDN regions exhibit temporal dynamics that closely resemble those observed during the resting state.

This remarkable consistency between the TDN and the DMN warrants thorough and extensive investigation, as it appears to relate to several important yet poorly understood questions in brain mapping—such as the precise cognitive function of fMRI deactivation and the nature of the brain's functional baseline. While a comprehensive discussion of this consistency is beyond the scope of the present paper, we will focus on its implications for the hypothesis that the DMN may serve as the brain's functional baseline (Gusnard & Raichle, 2001; Raichle, 2001), as this question is directly relevant to the central theme of this paper: brain-inspired AI modeling.

The hypothesis that the DMN serves as the brain's functional baseline is undoubtedly significant, yet it faces challenges from two main perspectives. First, the unconstrained nature of resting-state acquisitions poses a problem: treating "fixation" or "eyes-closed" conditions as a baseline is problematic, as there is no real control over cognitive processes during these so-called "resting" states. This lack of control undermines the validity of cognitive interpretations. Second, there is a lack of theoretical justification for considering the resting state as a fundamental baseline of brain function. Currently, no cognitive theory supports the notion that the resting state is fundamentally distinct from other brain states or that it should be afforded a privileged status as a functional baseline.

With respect to these two main challenges, it is noteworthy that the observed consistency between the TDN and the DMN may contribute to a potential resolution. On one hand, the topological discrimination task used to define the TDN, in contrast to "resting" conditions, offers the advantage of definite operational variables and well-controlled paradigms. These variables—including topological invariants—are mathematically well-defined as invariants under transformation groups, in the spirit of the Erlangen Program. The paradigms effectively control for various confounds, particularly nontopological ones, and make it highly unlikely that the TDN arises from responses to specific stimuli or figures rather than from the processing of topological properties. This strength of the TDN, in both its experimental variables and paradigms, suggests a solution to the unconstrained nature of resting-state acquisitions.

On the other hand, the "Global-first" principle seems to offer a solution to the lack of theoretical justification faced by the resting state of the DMN. The "Global-first" principle posits that perceptual organization based on global topological invariants serves as the starting point of cognitive processes. In particular, with respect to both time dependence and logical dependence, global topological properties—based on spatial and/or temporal proximity—take priority over local geometric and other featural properties. Thus, the "Global-first" principle highlights that topological perception is fundamentally distinct from local geometric and physical features, and should be afforded a privileged status as a functional baseline.

In short, the consistency between the TDN and the DMN supports the hypothesis that the DMN serves almost as the brain's functional baseline, but specifically from the perspective that the default, rather than the resting, TDN—like the DMN—could define this baseline.

*Epilepsy TDN Mapping: A Turning Point in Distinguishing the TDN from the DMN*

On one hand, this interesting and seemly important analysis on this consistency attracted us to make effort to search continuously for more systematic and robust evidence supporting this consistency by multiple behavioral paradigms, which performed at different field strengths (3T and 7T MRI scanner) as well as different types of signals measured (BOLD and CBF).

On other hand, despite this high degree of consistency, the fundamental difference between the TDN and the DMN in their modes of identification—task-evoked fMRI deactivations (particularly in topological discrimination tasks) versus resting-state fMRI synchrony (in, typically, "fixation" or "eye-closed resting")—still leaves us wondering which better informs the brain's functional baseline. Nevertheless, their remarkable and intriguing consistency in both spatial distribution and temporal dynamics has made differentiation between the TDN and the DMN difficult, particularly when analyses are limited to healthy subjects.

Surprisingly, our neuro-patient TDN scanning—particularly in cases of epilepsy associated with hippocampal sclerosis—led to a turning point in efforts to distinguish the TDN from the DMN (Note [158]). The results were striking: although our scanning revealed that approximately 92% of the epilepsy patients we tested exhibited standard DMN hubs, over 90% of these patients did not show significant TDN deactivation. This finding was robust not only in group analyses but also at the single-case level. Notably, these epilepsy patients performed the topological discrimination task quite well with no significant difference compared to healthy subjects. This behavioral result suggests that the absence of TDN deactivation was not due to an impairment in topological discrimination *per se*, but rather to the underlying pathological mechanisms of epilepsy.

This neuropsychological differentiation thereby elucidates that, going beyond the apparent consistency of the TDN and the DMN, the mechanism of the TDN intrinsically differs from that of the DMN. The applicability of the TDN, in contrast to the DMN, has been extended to the possible diagnosis of epilepsy, indicating that the TDN may reveal deeper mechanisms underlying brain function in both healthy individuals and neuro-patients. In this regard, the TDN may better inform the functional baseline of the brain.

More generally, while positive BOLD activation is well-established and typically associated with heightened neural activity, the intrinsic cognitive function of DMN deactivation—and its profound implications for understanding brain function—remains a critical and largely unresolved issue in brain mapping and cognitive science. Our discovery that TDN deactivation is consistently and systematically modulated by geometric invariants reveals its intrinsic cognitive role: processing the geometric invariants of the Erlangen Program as cognitive primitives. From the perspective of the "Global-first" principle, this function of TDN deactivation is fundamental for cognition—not merely suppressing non-essential regions to enhance efficiency and focus, as is often assumed. Processing geometric invariants at the scale of the whole brain, in the spirit of the Erlangen Program, likely has broad and far-reaching implications for various cognitive models and debates, such as picture-like versus propositional representations, metaphor mapping in semantics, mental rotation and imagery, and the modularity of mind. A detailed discussion of these impacts is beyond the scope of this paper, but we briefly highlight a key point. In current cognitive and AI models, activation—typically reflected by the strength of neural connections and firing rates—is commonly linked to heightened cognitive or computational functions, whereas deactivation is often overlooked or insufficiently considered. However, for brain-like or brain-inspired AI, alongside the inspiration discussed above regarding the limitation of connectionism—that while connections are important, the inner structure of neurons is too—the "Erlangen Program" for TDN deactivation offers another fundamental inspiration: while activation is fundamental, deactivation is too.

As suggested in our discussion above on the "Erlangen Program" for AI, on the "Erlangen Program" for visual cognition, and on the "Erlangen Program" for brain mapping, the Erlangen Program may particularly serve as a scientifically well-defined and empirically and formally interactive meeting point, bringing cognitive scientists and computational researchers together to work toward understanding RCC.

To sum up briefly, RCC, as illustrated by its four rich categories discussed above, is multifaceted, multidimensional, multilevel, interdisciplinary, and transdisciplinary. To capture RCC's initial core, the "Global-first" approach specifically focuses RCC on the contrast between "Global-first" cognition and local-first computation. This specification renders RCC a scientifically well-defined and methodologically approachable question, making the investigation of RCC not only feasible but also systematic and productive. In turn, from the perspective of brain and cognitive science, the "Global-first" principle may play an irreplaceable role in advancing toward brain-inspired AI or AGI by addressing cognitive primitives, cognitive neuroanatomy, cognitive emergence, and mathematical branches adapted to describe cognition.

The keyword "first" in the "Global-first" principle resonates with the "first" in the proverb "The first step is the hardest, yet it's not even hard". Despite centuries of effort in cognitive science and AI—and despite the historically unprecedented boom in computing power and AI modeling—we still do not know exactly "Where to begin?", or whether to go global-to-local or local-to-global. The "Global-first" principle offers a specific starting point to explore RCC: "Global-first" cognition versus local-first computation. This may serve as a breakthrough in our long journey toward understanding RCC.

**Acknowledgments**

This work was supported by the Chinese Academy of Sciences 0-1 Original Innovation Program (No. ZDBS-LY-SM028), the Hainan Academician Innovation Platform Scientific Research Project (No. PSPTZX202203), and the 1.3.5 Project for Disciplines of Excellence from West China Hospital, Sichuan University (No. ZYGD22003). Part of this work was conducted during L. Chen's adjunct professorship at West China Hospital, Sichuan University.

**Note**

Experimental evidence supporting the "Global-first" principle and the "Global-first" topological definition of the primitives of cognition is exemplified below, along with a brief introduction to each topic.

1. The highest visual sensitivity and discriminability are observed in response to topological

differences [1-2]. 2. A hierarchical relationship between Gestalt principles has been discovered: Proximity, characterized by tolerance connectivity in the homology of tolerance spaces, is the most fundamental and takes precedence over similarity; and similarity based on topological properties takes precedence over similarity based on local geometric properties [3-8; 4 (pp. 276-283); 7 (pp. 595-604);]. 3. A formally grounded topological definition of the figure-background relationship has been established and shown to be cognitively real and general [4 (p. 252); 7 (pp. 560-570); 9-10]. 4. A perceptual hierarchy of the relative salience of geometric invariants was found to be remarkably consistent with Klein's Erlangen Program [1; 2; 7; 11-13]. 5. Connectivity and holes, rather than terminators, give rise to immediate and effortless texture segregation [4; 14-18]. 6. A formal definition of perceptual objects has been established, utilizing topological invariants - connectivity, holes, and the inside/outside relationship - all of which are based on spatial and/or temporal proximity, and has been demonstrated to be both cognitively real and general [7 (pp. 566-570); 19 (Chs. 3 and 4); 20]. 7. The temporal window of subjective presence or simultaneity can be characterized by tolerance connectivity in time [21-30]. 8. Configural superiority effects arise from the dominance of topological features over local features [4; 7; 14; 31-32]. 9. The object-superiority effect is attributed to organization based on topological properties, including connectivity, holes, and inside/outside relations [19 (Ch. 4); 20; 31; 33-42]; 10. Global precedence essentially reflects the primacy of proximity, specifically tolerance connectivity, over similarity [5-6; 43-46]. 11. The topological type of similarity and response assignment overwhelms the formidable Redundant-Target Effect (RTE) [35]. 12. The topological preference in apparent motion demonstrates that topological invariants function as correspondence tokens [47-56]. 13. The guidance of visual search is predominantly influenced by topological properties [57-64]. 14. Topological changes disrupted object continuity in attentive tracking, suggesting the emergence of new objects [65]. 15. Topology-defined objects leave no space for space-based models in precuing attention [19 (Ch. 4); 66]. 16. Topological changes have been reported to cause attentional capture in a stimulus-driven manner, with a strength equal to that of abrupt onsets [67 (Ch. 3); 68; 69 (Chs. 3 and 4)]. 17. Topological changes in space and/or time between neighboring items in RSVP consistently triggered the attentional blink [70-72], even under conditions with just a single target [72] rather than a first target followed by a second target, and in two spatially separate RSVPs [71 (Ch. 3)]. 18. Topological difference between target and flankers alleviates crowding effect [73 (Ch. 5); 74-75]. 19. Eye movement was guided by topological changes resulting the emergence of new objects [76-78]. 20. Topological dominance was discovered in peripheral vision [46; 73 (Chs. 3 and 4); 79-80]. 21. Topological variations significantly contribute to the effects of visual masking [81; 82 (Chs. 3 and 4); 83]. 22. Illusory conjunctions of holes and the inside/outside relation were consistently found [84]. 23. The topological superiority effects were demonstrated in 3D form perception [85-87]. 24.

Working memory capacity is effectively measured using topology-defined units [88-93]. 25. No perceptual learning was found in topological discrimination tasks, yet unidirectional transferability was observed from Euclidean to affine, and then to projective properties [94 (Ch. 2); 95 (Chs. 2 and 3); 96 (Chs. 3 and 4); 97]. 26. Topological invariance dominated in the left hemispheres [98-99]. 27. The numbers counted in numerosity perception are fundamentally determined by topology-defined units in space and time [100-117]. 28. The underlying units of perceived animacy are discrete objects defined by connectivity [118]. 29. A topological approach to the spatial aspects of olfaction [119]. 30. The interaction between fear and topological perception was particularly found under well-controlled conditions: fear interfered specifically with topological (rather than local geometrical) perception, and only fear (as opposed to other basic emotions) interfered with topological perception [120-123]. 31. In binocular rivalry, the stimulus strength defined by topology dominated the transfer from preconsciousness to consciousness [124-126]. 32. A topological explanation of visual illusions and magical experiences [127-128]. 33. The topological nature of "the integrative self" was revealed by the finding that topological differences between self-associated forms presented in a priming task eliminated their self-reference effect [129]. 34. Infants, including those only hours old, demonstrated discriminability and sensitivity to topological differences [130-136]. 35. Children's form categorization, spatial representation, and numerical perception are all based on topological properties [137-142]. 36. Global topological structure in small brain of honey bees [143-145]. 37. Pigeons discriminated shapes based on topological features [146-148]. 38. In zebrafish, apparent motion activated the tectum in a manner similar to Klein's hierarchy of geometries, as measured with OKR and calcium imaging [149-153; 149 (Ch. 2); 153 (Ch. 3)]. 39. Dogs underestimated the quantity of connected items [154]. 40. Topological selectivity of neurons in the inferior temporal cortex of the monkey [8; 155-156]. 41. Topological perception originates in ipRGCs in mouse: behavioral and neurogenetic evidence [157]. 42. Epilepsy TDN mapping differentiated the TDN from the DMN, defined by either task-induced deactivation or resting-state functional connectivity, and indicates that the TDN better informs the functional baseline of the brain [158]. 43. Hemispatial neglect was exacerbated by the topological invariant of connectedness [159]. 44. Performance of topological perception in individuals with myopia [160-161]. 45. The global-first topological approach for Alzheimer's disease severity staging [162]. 46. Proximity grouping was significantly poorer in individuals with Williams syndrome [163]. 47. Autistic individuals exhibited less grouping-induced bias in numerosity judgments [164]. 48. Impairment in proximity organization and the subcortical visual pathway may contribute to the development of schizophrenia [165-166]. 49. Blight sight: Topological sight in blind field [167]. 50. Notably, no senescence of topological discriminability was discovered in age-related decline measurements of visual cognition, whereas the degrees of progressive decline with aging systematically increased according to the stabilities of local geometries in the Erlangen Program [168]. 51. The subcortical pathway is responsible for the fast processing of topological perception [120 (Ch. 5); 169-172; 170 (Ch. 3) ]. 52. fMRI scanning revealed that apparent motion produced activation in the anterior temporal lobe; notably, this activation was monotonically correlated with the form stability of the figures, in a manner similar to the Erlangen Program [173]. 53. Default but not rest: The topological discrimination network (TDN) defines the functional baseline of the brain [174]. 54. Computational analysis and modeling of topological perception [175-182]. 55. "Global-first" topological perception: a current century cloud over computational approaches based on local features [4 (pp. 629-633); 7; 14; 157; 175; 183]. 56. Closure defined a new perceptual category [184-185]. 57. A topological approach to psychophysiological changes associated with Chinese calligraphy [186]. 58. Rapid processing of topological relations between objects and concept acquisition [187-194]. 59. Event-related-potential studies of topological perception [195-197].

[Doctoral dissertation, University of Chinese Academy of Sciences].

[121] Meng, Q., Qian, W., Ren, P., Liu, N., Zhou, K., Ma, Y., & Chen, L. (2012). Interference between fear emotion and topological perception and its neural correlation in amygdala. *Journal of Vision*, **12**(9), 1307-1307. https://doi.org/10.1167/12.9.1307

[122] Xu, L., Su, H., Xie, X., Yan, P., Li, J., & Zheng, X. (2018). The topological properties of stimuli influence fear generalization and extinction in humans. *Frontiers in Psychology*, **9**, 409. https://doi.org/10.3389/fpsyg.2018.00409

[123] Huang, Y., Li, L., Dong, K., Tang, H., Yang, Q., Jia, X., Liao, Y., Wang, W., Ren, Z., Chen, L., & Wang, L. (2020). Topological shape changes weaken the innate defensive response to visual threat in mice. *Neuroscience Bulletin*, **36**(4), 427-431. https://doi.org/10.1007/s12264-019-00454-w

[124] Chen, Y. (2011). *Global first in implicit perception - A study using continuous flash suppression paradigm.* [Doctoral dissertation, University of Chinese Academy of Sciences].

[125] Meng, Q., Cui, D., Zhou, K., Chen, L., & Ma, Y. (2012). Advantage of hole stimulus in rivalry competition. *PLoS ONE*, **7**(3), e33053. https://doi.org/10.1371/journal.pone.0033053

[126] Chen, Y., Zhou, T., & Chen, L. (2014). Implicit perception of "holes" under continuous flash suppression. *Progress in Biochemistry and Biophysics*, **41**(6), 551-557. https://doi.org/10.3724/Sp.J.1206.2013.00463

[127] Ekroll, V. (2019). Illusions of imagery and magical experiences. *i-Perception*, **10**(4), 2041669519865284. https://doi.org/10.1177/2041669519865284

[128] Nemati, F. (2022). The suitability of topology for the investigation of geometric-perceptual phenomena. *Phenomenology and the Cognitive Sciences*, **24**(2), 395-410. https://doi.org/10.1007/s11097-022-09857-z

[129] Xi, H. (2020). *The relationship between self as object and topological perception: The topological nature of self.* [Doctoral dissertation, University of Chinese Academy of Sciences].

[130] Colombo, J., Laurie, C., Martelli, T., & Hartig, B. (1984). Stimulus context and infant orientation discrimination. *Journal of Experimental Child Psychology*, **37**(3), 576-586. https://doi.org/10.1016/0022-0965(84)90077-8

[131] Farran, E. K., Brown, J. H., Cole, V. L., Houston-Price, C., & Karmiloff-Smith, A. (2008). A longitudinal study of perceptual grouping by proximity, luminance and shape in infants at two, four and six months. *International Journal of Developmental Science*, **2**(4), 353-369. https://doi.org/10.3233/dev-2008-2402

[132] Turati, C., Simion, F., & Zanon, L. (2010). Newborns' perceptual categorization for closed and open geometric forms. *Infancy*, **4**(3), 309-325. https://doi.org/10.1207/s15327078in0403_01

[133] Chien, S. H. L., & Lin, Y. L. (2011). With or without a hole: Young infant's sensitivity for topological. *i-Perception*, **2**(4), 317-317. https://doi.org/10.1068/ic317

[134] Chien, S. H. L., Lin, Y. L., Qian, W., Zhou, K., Lin, M. K., & Hsu, H. Y. (2012). With or without a hole: Young infants' sensitivity for topological versus geometric property. *Perception*, **41**(3), 305-318. https://doi.org/10.1068/p7031

[135] Kibbe, M. M., & Leslie, A. M. (2016). The ring that does not bind: Topological class in
34